\documentclass[a4paper,11pt]{article}
\usepackage[left=3.17cm,right=3.17cm,top=2.54cm,
headheight=0.5cm,headsep=0.54cm,bottom=2.54cm,footskip=0.79cm
]{geometry}

\usepackage{amsmath,amssymb}
\usepackage[pdfstartview=FitH]{hyperref}

\numberwithin{equation}{section}

\begin{document}

\title{Deformed squeezed states in noncommutative phase space
\thanks{This project supported by National Natural
Science Foundation of China under Grant 10675106.}}

\author{Bingsheng Lin\thanks{Corresponding author.} , Sicong Jing\\
\textit{\small{Department of Modern Physics, University of Science
and Technology of China}}\\
\textit{\small{Hefei, Anhui 230026, China}}
}
\date{14 January 2008}
\maketitle
\footnotetext[1]{\textit{E-mail addresses:} gdjylbs@mail.ustc.edu.cn (B. Lin), sjing@ustc.edu.cn (S. Jing).}

\begin{abstract}
\noindent
A deformed boson algebra is naturally introduced from studying
quantum mechanics on noncommutative phase space in which both
positions and momenta are noncommuting each other. Based on this
algebra, corresponding intrinsic noncommutative coherent and
squeezed state representations are constructed, and variances of
single- and two-mode quadrature operators on these states are
evaluated. The result indicates that in order to maintain
Heisenberg's uncertainty relations, a restriction between the
noncommutative parameters is required.\\
\ \\
\textit{PACS:} 02.40.Gh; 03.65.Ta; 03.65.Fd\\
\ \\
\textit{Keywords:} Noncommutative phase space; Deformed boson algebra; Squeezed
state; Heisenberg uncertainty relation
\end{abstract}

\section{Introduction}\label{sec1}
Recently there has been much interest in the study of physics in
noncommutative space, mainly due to the noncommutative space is
necessary for describing low-energy effective theory of a D-brane
with a B-field background, but also the noncommutativity of
space-time could play an important role in quantum gravity. Besides
a lot of papers contributed to the study of the noncommutative field
theory \cite{s1,s2}, there are many works devoted to the study
of various aspects of quantum mechanics on the noncommutative space
with usual (commutative) time coordinate \cite{s3}-\cite{s5}.
Although in string theory only the coordinate space exhibits a
noncommutative structure, some authors have studied models in which
a noncommutative geometry \cite{s6} defines the whole phase space
\cite{s7, s8}. Noncommutativity between momenta arises naturally
as a consequence of noncommutativity between coordinates, as momenta
are defined to be the partial derivatives of the action with respect
to the noncommutative coordinates \cite{s9}.

The usual method to study the noncommutative quantum mechanics
(NCQM) is using the Seiberg-Witten map to change a problem of NCQM
into a corresponding problem of quantum mechanics on the commutative
space. In the case of only the coordinate space is noncommutative,
this method is consistent with the Weyl-Moyal correspondence which
amounts to replacing the usual product by a star product in
noncommutative space. This method, however, does not always work,
for example, when both coordinates and momenta are noncommutative,
i.e., on a noncommutative phase space, although using the
Seiberg-Witten map one can write a Hamiltonian of NCQM in terms of
the ordinary commutative coordinates and momenta variables, one has
no way to get a well-defined Schr\"{o}dinger equation which is
consistent with the corresponding star product. Therefore, it is
necessary to develop other new method to solve the quantum
mechanical problems directly in the noncommutative phase space.

Our another motivation is to build some effective representations in
NCQM. Since the noncommutativity between different coordinate (or
momentum) component operators, there are no simultaneous eigenstates
of these coordinate (or momentum) operators, therefore no exist
ordinary coordinate (or momentum) representations in NCQM. However,
in order to formulate quantum mechanics on the noncommutative space
so that some dynamical problems can be solved, we do need some
appropriate representations.

In this Letter we study how to construct noncommutative coherent and
squeezed state representations for NCQM. We show that the coherent
and squeezed state representations can be constructed equally well
not only from the ordinary boson algebraic relations but also from a
kind of deformed boson commutation relations which can be derived
from a four-dimensional noncommutative phase space defined by the
noncommutative parameters $\mu$ and $\nu$ in the following way
\begin{align}\label{ncxp}
[\hat{x},\,\hat{y}]&=\emph{i}\,\mu\,,&
[\hat{p}_x,\,\hat{p}_y]&=\emph{i}\,\nu\,,\nonumber\\
[\hat{x},\,\hat{p}_x]&=[\hat{y},\,\hat{p}_y]=\emph{i}\,\hbar\,,&
[\hat{x},\, \hat{p}_y]&=[\hat{y},\,\hat{p}_x]=0\,,
\end{align}
where without loss of generality we have taken $\mu$, $\nu$ as
positive real parameters. A deformed boson algebra was derived from
such kind of deformed Heisenberg-Weyl algebra (\ref{ncxp}) under an
assumption of maintaining Bose-Einstein statistics \cite{s10}, (this
assumption is equivalent to propose a direct proportionality between
the noncommutative parameters $\mu$ and $\nu$), and some authors
further investigated variances of the deformed single- and two-mode
quadrature operators in the noncommutative phase space
\cite{s11}-\cite{s13}. However, recently Bertolami and Rosa argued
that there is no strong argument supporting such a direct
proportionality relation \cite{s14}. In this Letter we derive the same
deformed boson algebra (Eq.(\ref{ncab}), see below) without using any
additional assumption. All our analysis and calculation are
performed directly in the noncommutative phase space and not by
virtue of corresponding variables in the commutative space which
appear via some kind of Seiberg-Witten map.

The work is organized as follows. In next section we use a simple
method to derive the deformed boson algebra from the commutation
relations in Eq.(\ref{ncxp}). Based on this deformed boson algebra
we introduce two-mode displacement and squeeze operators in the
noncommutative phase space and show that the two-mode squeeze
operator induces a generalized Bogoliubov's transformation. In
section \ref{sec3} two-mode coherent and squeezed states are constructed on
the noncommutative phase space and some basic properties of these
states (such as inner product and overcompleteness) are given, which
enable them to be effective representations. In section \ref{sec4}, variances
of the deformed single- and two-mode quadrature operators in the
noncommutative phase space are evaluated, and a constrain relation
between the noncommutative parameters $\mu$ and $\nu$ is derived
from the Heisenberg uncertainty relations. The last section devotes
to some discussion and comment.
\newpage

\section{Deformed bosonic realization of noncommutative phase space}
\label{sec2}
We start from Eq.(\ref{ncxp}) on the noncommutative phase space and
introduce following deformed boson operators
\begin{equation}\label{ab}
\hat{a}=\frac{1}{\sqrt{2\hbar}}\left( \sqrt[4]{\frac{\nu}{\mu}}\,
\hat{x} + \emph{i}\,\sqrt[4]{\frac{\mu}{\nu}} \,\hat{p}_x \right),
~~~~~~ \hat{b}=\frac{1}{\sqrt{2\hbar}}\left(
\sqrt[4]{\frac{\nu}{\mu}}\, \hat{y} +
\emph{i}\,\sqrt[4]{\frac{\mu}{\nu}}\, \hat{p}_y \right)
\end{equation}
and their Hermitian conjugate operators $\hat{a}^{\dag}$,
$\hat{b}^{\dag}$, which satisfy commutation relations
\begin{equation}\label{ncab}
[\hat{a},\,\hat{a}^{\dag}]=[\hat{b},\,\hat{b}^{\dag}]=1,~~~~
[\hat{a},\,\hat{b}]=[\hat{a}^{\dag},\,\hat{b}^{\dag}]=0,~~~~
[\hat{a},\,\hat{b}^{\dag}]=-[\hat{b},\,\hat{a}^{\dag}]=\emph{i}\,\theta,
\end{equation}
where $\theta=\sqrt{\mu \nu}/\hbar$\,. Here we would like to
emphasize that the expressions in (\ref{ab}) only work for the case of
$\mu$ and $\nu$ both nonzero, and this is just the situation we are
considering. The algebraic relations in Eq.(\ref{ncab}) are exactly
the same in form as the deformed boson algebra in \cite{s10}. To our
knowledge, this type of deformed boson commutation relations also
appeared in early work by Caves and Schumaker \cite{s15}. Their
quadrature-phase amplitudes satisfy the similar commutation
relations as Eq.(\ref{ncab}). When $\theta=0$, Eq.(\ref{ncab})
reduces to ordinary boson algebra.

An unitary two-mode displacement operator in the noncommutative
phase space may be defined as
\begin{equation}\label{tmdo}
D(\alpha, \beta)=\exp(\alpha \hat{a}^{\dagger} + \beta
\hat{b}^{\dagger} - \alpha^{\ast} \hat{a} - \beta^{\ast} \hat{b}),
\end{equation}
with the important property
\begin{equation}
D(\alpha,\beta)^{\dagger}\,\hat{a}\,D(\alpha,\beta)
=\hat{a}+\alpha+\emph{i}\,\theta\beta,~~~~
D(\alpha,\beta)^{\dagger}\,\hat{b}\,D(\alpha,\beta)
=\hat{b}+\beta-\emph{i}\,\theta\alpha.
\end{equation}
Now we introduce a two-mode squeeze operator in the noncommutative
phase space
\begin{equation}\label{tmso}
\hat{S}(z)=\exp ( z^{\ast} \hat{a}\hat{b}
-z\hat{a}^{\dagger}\hat{b}^{\dagger}).
\end{equation}
Inspection of the above operator shows that $\hat{S}(z)^{\dagger}=
\hat{S}(z)^{-1}=\hat{S}(-z)$. Using Eq.(\ref{ncab}) after straightforward
calculation one can derive the following transformation
$(z=r\,e^{\emph{i}\,\varphi})$
\begin{eqnarray}
\hat{S}(z) \hat{a} \hat{S}(z)^{\dagger}
&=&(\cosh r\,\cosh r\theta)\hat{a}
+i(\sinh r\,\sinh r\theta)\hat{b}\nonumber\\
&&+e^{i\,\varphi}(\sinh r\,\cosh r\theta)\hat{b}^{\dag}
 + ie^{i\,\varphi}(\cosh r\,\sinh r\theta)\hat{a}^{\dag},\nonumber\\
\hat{S}(z) \hat{b} \hat{S}(z)^{\dagger}
&=&(\cosh r\,\cosh r\theta)\hat{b}
-i(\sinh r\,\sinh r\theta)\hat{a}\nonumber\\
&&+ e^{i\,\varphi}(\sinh r\,\cosh r\theta)\hat{a}^{\dag}
 - ie^{i\,\varphi}(\cosh r\,\sinh r\theta)\hat{b}^{\dag},
\end{eqnarray}
which clearly generalizes well-known Bogoliubov's transformation
\cite{s16} in commutative space to the noncommutative phase space,
and of course, when $\theta=0$, it reduces to the ordinary case. So
we will call it generalized or deformed \mbox{Bogoliubov}'s transformation.
In fact, a $q$-deformed Bogoliubov's transformation for a pair of
$q$-oscillators has been considered by A. Zhedanov \cite{s17}, who
has also shown that the $q$-deformed Bogoliubov's transformation is
nonlinear and is related to Kravchuk's $q$-polynomials. For the
later convenience, we denote the squeezed deformed boson operators
$\hat{a}$ and $\hat{b}$ as
\begin{equation}
\hat{A} = \hat{S}(z) \,\hat{a}\, \hat{S}(z)^{\dagger},~~~~~~ \hat{B} =
\hat{S}(z) \,\hat{b}\, \hat{S}(z)^{\dagger}.
\end{equation}
Of course, the operators $\hat{A}$, $\hat{B}$ satisfy the same
commutation relations as $\hat{a}$ and $\hat{b}$ do (Eq.(\ref{ncab}))
because they are related each other by an unitary squeeze
transformation.

\section{Two-mode coherent and squeezed states in noncommutative phase space}
\label{sec3}
Eq.(\ref{ncab}) shows that the operators $\hat{a}$ and $\hat{b}$ are
commuting each other and they have simultaneous eigenstates, i.e.,
coherent states. Starting from a vacuum state $|0 \rangle$, which is
normalized and destroyed by the deformed boson operators $\hat{a}$
and $\hat{b}$, and using the two-mode displacement operator
(\ref{tmdo}), one can construct a two-mode coherent state in the
noncommutative phase space
\begin{equation}
|\alpha,\beta \rangle =D(\alpha,\beta)|0 \rangle,
\end{equation}
which satisfies
\begin{equation}
\hat{a}|\alpha,\beta \rangle= (\alpha + \emph{i}\,\theta \beta)
|\alpha,\beta \rangle, ~~~~\hat{b}|\alpha,\beta \rangle= (\beta -
\emph{i}\,\theta \alpha) |\alpha,\beta \rangle,
\end{equation}
respectively. The inner product of two such coherent states is
easily to get
\begin{eqnarray}\label{abab}
\langle \alpha',\beta'|\alpha,\beta \rangle
 &=& \exp \Big\{
-\frac{1}{2}\big(|\alpha|^2 +|\beta|^2 +|\alpha'|^2 +|\beta'|^2\big)
 +\alpha'^{\ast} \alpha + \beta'^{\ast} \beta \nonumber\\
&&+ \frac{\emph{i}\,\theta}{2}(\beta^{\ast}\alpha -
\alpha^{\ast}\beta +\beta'^{\ast}\alpha'-\alpha'^{\ast}\beta')
+\emph{i}\,\theta (\alpha'^{\ast}\beta - \beta'^{\ast}\alpha)
\Big\},
\end{eqnarray}
which means that the two-mode coherent states are normalized but not
orthogonal to each other, and besides, they are over-complete. The
corresponding resolution of the identity is
\begin{equation}\label{compl}
(1-\theta^2) \int \frac{d^2 \alpha\,d^2 \beta}{\pi^2} |\alpha,\beta
\rangle\,\langle \alpha,\beta |=1,
\end{equation}
where the integral is over two complex planes, and
$d^2 \alpha =d{\rm Re}(\alpha)\,d{\rm Im}(\alpha)$,
$d^2\beta = d{\rm Re}(\beta)\,d{\rm Im}(\beta)$.

In the noncommutative phase space, a two-mode squeezed state
$|\alpha,\beta;z \rangle$ can be defined as
\begin{equation}
|\alpha,\beta;z \rangle = \hat{S}(z)|\alpha,\beta \rangle,
\end{equation}
which is simultaneous eigenstate of the squeezed boson operators
$\hat{A}$ and $\hat{B}$,
\begin{equation}\label{abvalue}
\hat{A}|\alpha,\beta;z \rangle
= (\alpha + \emph{i}\,\theta \beta)|\alpha,\beta;z \rangle,~~~~
\hat{B}|\alpha,\beta;z \rangle
= (\beta- \emph{i}\,\theta \alpha) |\alpha,\beta;z \rangle.
\end{equation}
Multiplying Eq.(\ref{compl}) by $\hat{S}(z)$ from the left-side and by
$\hat{S}^{\dagger}(z)$ from the right-side, one has
\begin{equation}
(1-\theta^2) \int \frac{d^2 \alpha\,d^2 \beta}{\pi^2}
|\alpha,\beta;z \rangle\,\langle \alpha,\beta;z |=1,
\end{equation}
which means that the two-mode squeezed states also satisfy the
resolution of identity. Thus we have coherent representation and
squeezed representation in the noncommutative phase space, and any
state in this space can be expanded in terms of $|\alpha,\beta
\rangle$ or $|\alpha,\beta;z \rangle$.

Using the algebra (\ref{ncab}) one can straightforwardly evaluate the
wavefunction of the two-mode squeezed state $|\alpha,\beta;z
\rangle$ in the two-mode coherent state representation, i.e.,
\begin{equation}\label{ababz}
\begin{split}
&\langle \alpha',\beta'|\alpha,\beta;z \rangle = \Big( \cosh (
r(1+\theta) )\cosh ( r(1-\theta) ) \Big)^{-1/2}\\
&\times \exp \Big( -\frac{1}{2} \big(|\alpha|^2 +|\beta|^2 +
|\alpha'|^2 + |\beta'|^2 + \emph{i}\,\theta (\alpha^{\ast}\beta -
\beta^{\ast}\alpha + \alpha'^{\ast}\beta' - \beta'^{\ast}\alpha' )
\big) \Big)\\
&\times\exp \bigg\{ \frac{(1+\theta)(\alpha'^{\ast}\alpha
+\beta'^{\ast}\beta +\emph{i}\,\alpha'^{\ast}\beta -\emph{i}\,
\beta'^{\ast}\alpha)}{2\cosh ( r(1+\theta) )} +
 \frac{(1-\theta)(\alpha'^{\ast}\alpha +\beta'^{\ast}\beta
-\emph{i}\,\alpha'^{\ast}\beta +\emph{i}\,
\beta'^{\ast}\alpha)}{2\cosh ( r(1-\theta) )}
\bigg\}\\
&\times\exp \bigg\{ -\emph{i}\,e^{-\emph{i}\,\varphi} \Big(
\frac{1+\theta}{4}\tanh (r(1+\theta))(\alpha +\emph{i}\,\beta )^2 -
\frac{1-\theta}{4}\tanh (r(1-\theta))(\alpha -\emph{i}\,\beta )^2
\Big)\\
&- \emph{i}\,e^{\emph{i}\,\varphi} \Big( \frac{1+\theta}{4}\tanh
(r(1+\theta))(\alpha'^{\ast} -\emph{i}\,\beta'^{\ast} )^2 -
\frac{1-\theta}{4}\tanh (r(1-\theta))(\alpha'^{\ast}
+\emph{i}\,\beta'^{\ast} )^2 \Big) \bigg\}.
\end{split}
\end{equation}
Obviously when $z=0$, the above expression will reduce to
Eq.(\ref{abab}). Having (\ref{ababz}) and using (\ref{compl}), one can evaluate
another important inner product of two arbitrary such squeezed
states $\langle \alpha',\beta';z'|\alpha,\beta;z \rangle$.

\section{Heisenberg uncertainty relations}\label{sec4}
Obviously, the commutation relations (\ref{ncxp}) lead to the following
uncertainty relations
\begin{equation}\label{uncr}
\triangle \hat{x} \triangle \hat{y} \geqslant \frac{\mu}{2},~~~~
\triangle \hat{p}_x \triangle \hat{p}_y \geqslant \frac{\nu}{2},~~~~
\triangle \hat{x} \triangle \hat{p}_x \geqslant \frac{\hbar}{2},~~~~
\triangle \hat{y} \triangle \hat{p}_y \geqslant \frac{\hbar}{2}.
\end{equation}
The first two uncertainty relations show that measurements of
positions and momenta in both directions $x$ and $y$ are not
independent. Taking into account the fact that $\mu$ and $\nu$ have
dimensions of $(length)^2$ and $(momentum)^2$ respectively, then
$\sqrt \mu$ and $\sqrt \nu$ define fundamental scales of length and
momentum which characterize the minimum uncertainties possible to
achieve in measuring these quantities. One expects these fundamental
scales to be related to the scale of the underlying field theory
(possible the string scale), and thus to appear as small corrections
at the low-energy level or quantum mechanics.

Now one can evaluate the variances of the single-mode quadrature
operators $\hat{x}$ and $\hat{p}_x$ on the state $|\alpha,\beta;z
\rangle$ and get
\begin{eqnarray}\label{dx2}
(\triangle \hat{x})_z^2 &=& \langle \alpha,\beta;
z|\hat{x}^2|\alpha,\beta;z \rangle -  \langle \alpha,\beta;
z|\hat{x}|\alpha,\beta;z \rangle^2 \nonumber\\
&=& \frac{\hbar}{2} \sqrt{\frac{\mu}{\nu}} \Big( \cosh 2r (
\cosh 2r\theta +\sin \varphi \sinh 2r\theta ) \nonumber\\
&&~~~~ +\, \theta \sinh 2r ( \sinh 2r\theta + \sin \varphi \cosh
2r\theta) \Big),\nonumber\\
(\triangle \hat{p}_x)_z^2&=& \langle \alpha,\beta;
z|\hat{p}_x^2|\alpha,\beta;z \rangle -  \langle \alpha,\beta;
z|\hat{p}_x|\alpha,\beta;z \rangle^2  \nonumber\\
& =& \frac{\hbar}{2} \sqrt{\frac{\nu}{\mu}} \Big( \cosh 2r (
\cosh 2r\theta - \sin \varphi \sinh 2r\theta ) \nonumber\\
&&~~~~ +\theta \sinh 2r ( \sinh 2r\theta - \sin \varphi \cosh
2r\theta) \Big),
\end{eqnarray}
where the subscript $z$ on the left-hand side of Eq.(\ref{dx2}) means the
variance is for the state $|\alpha,\beta;z \rangle$. Furthermore,
one has
\begin{equation}
\sqrt{\frac{\nu}{\mu}} (\triangle \hat{x})_z^2 =
\sqrt{\frac{\mu}{\nu}} (\triangle \hat{p}_y)_z^2\,, ~~~~
\sqrt{\frac{\nu}{\mu}} (\triangle \hat{y})_z^2 =
\sqrt{\frac{\mu}{\nu}} (\triangle \hat{p}_x)_z^2\,.
\end{equation}
These results show that the variances of the single-mode quadrature
operators are independent on the complex numbers $\alpha,\,\beta$ in
the squeezed state $|\alpha,\beta;z \rangle$. They are dependent
only on the squeezing parameter $z$ and the noncommutative parameter
$\theta$.

Eq.(\ref{dx2}) leads to
\begin{multline}
(\triangle \hat{x})_z^2(\triangle \hat{p}_x)_z^2
=\frac{\hbar^2}{4}\Big[\cosh^2 2r\,\big(\cosh^2
2r\theta-\sin^2\varphi\,\sinh^2 2r\theta\big)
+\frac{1}{2}\theta\cos^2\varphi\,\sinh 4r\,\sinh 4r\theta \\
+\,\theta^2\sinh^2 2r\,\big(\sinh^2
2r\theta-\sin^2\varphi\,\cosh^2 2r\theta\big)\Big].
\end{multline}
One will find that for $\varphi=|\pi|/2$\,, $(\triangle
\hat{x})_z^2(\triangle \hat{p}_x)_z^2$\, reaches its minimum,
\begin{eqnarray}
\min \big((\triangle \hat{x})_z^2(\triangle \hat{p}_x)_z^2\big)
&=&\frac{\hbar^2}{4}\big(1+(1-\theta^2)\sinh^2 2r\big)\nonumber\\
&=&\frac{\hbar^2}{4}\big(1+(1- \frac{\mu \nu}{\hbar^2})\sinh^2
2r\big).
\end{eqnarray}
From this expression we see that there exists a natural constraint
for the noncommutative parameters $\mu$ and $\nu$, i.e., only when
$\mu\nu \leqslant \hbar^2$ the Heisenberg uncertainty relation $\triangle
\hat{x} \triangle \hat{p}_x \geqslant \hbar/2$ is satisfied, and if $\mu
\nu = \hbar^2$, $\triangle \hat{x} \triangle \hat{p}_x=\hbar/2$. The
same is for $\hat{y}$ and $\hat{p}_y$.

Also for $\varphi=|\pi|/2$ one has
\begin{eqnarray}
\min \big((\triangle \hat{x})_z^2(\triangle \hat{y})_z^2\big)
&=&\frac{\hbar^2 \mu}{4\nu}
\big(1+(1- \frac{\mu \nu}{\hbar^2}) \sinh^2 2r \big),\nonumber\\
\min \big((\triangle \hat{p}_x)_z^2(\triangle\hat{p}_y)_z^2\big)
&=& \frac{\hbar^2 \nu}{4\mu}
\big(1+(1- \frac{\mu\nu}{\hbar^2}) \sinh^2 2r \big).
\end{eqnarray}
It is easy to find that in order to have the Heisenberg uncertainty
relations $\triangle \hat{x} \triangle \hat{y} \geqslant \mu/2$ and
$\triangle \hat{p}_x \triangle \hat{p}_y \geqslant \nu/2$, it also needs
$\mu \nu \leqslant \hbar^2$, and for $\mu \nu=\hbar^2$, the Heisenberg
uncertainty relations become equalities.

It is interesting to compare the variances of these single-mode
quadrature operators on the state $|\alpha,\beta;z \rangle$ with
ones on the coherent state $|\alpha,\beta \rangle$. One can simply
find that the variances of $\hat{x}$ on the state $|\alpha,\beta
\rangle$ is
\begin{equation}\label{cdx2}
(\triangle \hat{x})^2=\langle \alpha,\beta| \hat{x}^2 |\alpha,\beta
\rangle -\langle \alpha,\beta| \hat{x} |\alpha,\beta \rangle^2 =
\frac{\hbar}{2} \sqrt{\frac{\mu}{\nu}}\,,
\end{equation}
and similarly
\begin{equation}
(\triangle \hat{p}_x)^2=\frac{\hbar}{2}\sqrt{\frac{\nu}{\mu}}\,,~~~~
(\triangle \hat{y})^2= (\triangle\hat{x})^2\,, ~~~~
(\triangle \hat{p}_y)^2=(\triangle \hat{p}_x)^2\,.
\end{equation}
These relations show that the last two Heisenberg's relations in
(\ref{uncr}) reach their minimums on the coherent states, however, the
first two relations in (\ref{uncr}) work also only for $\mu \nu\leqslant
\hbar^2$. Furthermore, from (\ref{dx2}) and (\ref{cdx2}), one has
\begin{eqnarray}\label{dx2z}
(\triangle \hat{x})_z^2
&=& \Big( \cosh 2r ( \cosh 2r\theta +\sin
\varphi \sinh 2r\theta )\nonumber\\
&&~~+\, \theta \sinh 2r ( \sinh 2r\theta + \sin
\varphi \cosh 2r\theta) \Big) (\triangle \hat{x})^2 ,
\end{eqnarray}
and also
\begin{eqnarray}\label{dp2z}
(\triangle \hat{p}_x)_z^2
&=& \Big( \cosh 2r ( \cosh 2r\theta - \sin
\varphi \sinh 2r\theta )\nonumber\\
&&~~~+ \theta \sinh 2r ( \sinh 2r\theta
- \sin\varphi \cosh 2r\theta) \Big) (\triangle \hat{p}_x)^2.
\end{eqnarray}
If $\theta=0$, (\ref{dx2z}) and (\ref{dp2z}) give
\begin{equation}
(\triangle \hat{x})_z^2 > (\triangle \hat{x})^2\,,~~~~
(\triangle\hat{p}_x)_z^2 > (\triangle \hat{p}_x)^2
\end{equation}
for nonzero squeezing parameter $z$, which mean that on the squeezed
state $|\alpha,\beta;z \rangle$ the corresponding Heisenberg
uncertainty relation cannot reach its minimum. Only due to the
existence of nonzero $\theta$, the case
\begin{equation}
(\triangle \hat{x})_z^2 \leqslant (\triangle \hat{x})^2 , ~~~~~~~~
(\triangle \hat{p}_x)_z^2 \leqslant (\triangle \hat{p}_x)^2
\end{equation}
may occur for some ranges of the parameter $z$ and the two-mode
squeezed state $|\alpha,\beta;z \rangle$ may exhibit some squeezing
effects for these cases.

It is worth while to point out that one can also choose, instead of
(\ref{ab}), the following operators \footnote{Authors thank one of the
referees to point out this.}
\begin{eqnarray}
&&a=(\sqrt{2\hbar})^{-1} \Big( \kappa - \frac{\mu \nu}{4\kappa
\hbar^2} \Big)^{-1} \left( \kappa \hat{x} + \frac{\mu}{2\hbar}
\hat{p}_y + \emph{i}\,\big(-\frac{\nu}{2\kappa \hbar} \hat{y} +
\hat{p}_x \big) \right), \nonumber\\
&&b=(\sqrt{2\hbar})^{-1} \Big( \kappa - \frac{\mu \nu}{4\kappa
\hbar^2} \Big)^{-1} \left( \kappa \hat{y} - \frac{\mu}{2\hbar}
\hat{p}_x + \emph{i}\,\big(\frac{\nu}{2\kappa \hbar} \hat{x} +
\hat{p}_y \big) \right),
\end{eqnarray}
to construct a basis of coherent or squeezed states to be used for
calculation of the expectation values. It is easy to show that, when
$\kappa = \big(1+ \sqrt{1- \frac{\mu \nu}{\hbar^2}}\,\big)/2$, the operators
$a,\,b$ and their Hermitian conjugates $a^\dag,\,b^\dag$ satisfy
ordinary boson commutation relations
\begin{equation}\label{bcr}
[a,\,a^\dag]=[b,\,b^\dag]=1,~~~~~~[a,\,b]=[a,\,b^\dag]=0.
\end{equation}
Now, with these ordinary boson operators, one can equally well
construct a two-mode overcomplete basis of coherent states as well
as squeezed states. Of course, the basic expressions for these
states are the standard ones. Starting from these ordinary coherent
states or ordinary squeezed states, one can readily evaluate the
variances for the coordinate and momentum operators $\hat{x}$,
$\hat{y}$, $\hat{p}_x$, $\hat{p}_y$ and get the same restriction $\mu \nu
\leqslant \hbar^2$ only from the condition that the parameter $\kappa$
has to be real in order to be able to introduce the above operators
$a$ and $b$. Also the expressions in (\ref{bcr}) work for any
noncommutative parameters $\mu$ and $\nu$, no matter they are zero
or nonzero. Needless to say, there are many possible choice of such
kind of maps.

Now we turn to discuss the two-mode quadrature operators which are
defined as
\begin{eqnarray}
\hat{X}
&=&\frac{\hat{x}+\hat{y}}{2} =\sqrt{\frac{\hbar}{2}}
\sqrt[4]{\frac{\mu}{\nu}}\,
\frac{\hat{a}+\hat{b}+\hat{a}^{\dag}+\hat{b}^{\dag}}{2}\,,\nonumber\\
\hat{P}
&=& \frac{\hat{p}_x+\hat{p}_y}{2} = \sqrt{\frac{\hbar}{2}}
\sqrt[4]{\frac{\nu}{\mu}}\,
\frac{\hat{a}+\hat{b}-\hat{a}^{\dag}-\hat{b}^{\dag}}{2\emph{i}}
\end{eqnarray}
and satisfy the following commutation relation
$[\hat{X},\hat{P}]=\emph{i}\,\hbar /2$\,. Similarly to the case of
the single-mode quadrature operators, one may derive the variances
of $\hat{X}$ and $\hat{P}$ on the two-mode squeezed state
$|\alpha,\beta;z \rangle$
\begin{eqnarray}
(\triangle \hat{X})_z^2
&=& \frac{\hbar}{4} \sqrt{\frac{\mu}{\nu}}
\big( \cosh 2r\theta (\cosh 2r - \cos \varphi \sinh2r) \nonumber\\
&&~~~~~~~~+\theta\sinh 2r\theta (\sinh 2r
- \cos \varphi \cosh 2r) \big), \nonumber\\
(\triangle \hat{P})_z^2
&=& \frac{\hbar}{4} \sqrt{\frac{\nu}{\mu}}
\big( \cosh 2r\theta (\cosh 2r + \cos \varphi \sinh2r) \nonumber\\
&&~~~~~~~~+\theta\sinh 2r\theta (\sinh 2r
+ \cos \varphi \cosh 2r) \big),
\end{eqnarray}
which lead to
\begin{multline}\label{dxdp}
(\triangle \hat{X})_z^2(\triangle \hat{P})_z^2
=\frac{\hbar^2}{16}\Big\{ (\cosh^2 2r-\cos^2\varphi\,\sinh^2 2r)
+\frac{\theta}{2} \sin^2\varphi\,\sinh 4r\,\sinh 4r\theta\\
+\big[(\cosh^2 2r +\theta^2\sinh^2 2r) -\cos^2
\varphi\,(\theta^2\cosh^2 2r +\sinh^2 2r) \big]\sinh^2 2r \theta
\Big\}.
\end{multline}
When $\varphi=0$, (\ref{dxdp}) reaches its minimum
\begin{equation}
\min \big((\triangle \hat{X})_z^2(\triangle \hat{P})_z^2 \big)
=\frac{\hbar^2}{16}\big(1+(1-\theta^2)\sinh^2 2r\theta \big),
\end{equation}
which also shows that the Heisenberg uncertainty relation
$(\triangle \hat{X})_z(\triangle \hat{P})_z \geqslant \hbar/4$ requires
$\mu\nu \leqslant \hbar^2$ and only for $\mu\nu =\hbar^2$ the uncertainty
relation becomes an equality.

\section{Discussion and conclusion}\label{sec5}
In this Letter we introduce the deformed boson operators $\hat{a}$,
$\hat{b}$ in (\ref{ab}) and their Hermitian conjugate operators
$\hat{a}^{\dag}$, $\hat{b}^{\dag}$, which satisfy the deformed boson
algebra (\ref{ncab}). Based on the deformed boson algebra, we
construct deformed two-mode coherent states and squeezed states on
the noncommutative phase space and show their inner products and
overcompleteness properties, which enable these states form
corresponding effective representations in the noncommutative phase
space. Then we evaluate the variances of the single- and two-mode
quadrature operators on the two-mode squeezed states
$|\alpha,\beta;z \rangle$ and investigate corresponding Heisenberg
uncertainty relations. From these variances, we find that there
exists a constraint between the noncommutative parameters $\mu$ and
$\nu$, and when $\mu \nu= \hbar^2$, all the Heisenberg inequalities
become equalities, which is in agreement with \cite{s18}. Our
analysis and calculation are performed on the noncommutative phase
space straightforwardly, without depending on any variables on the
commutative space.

It should be pointed out that when considering a concrete
Hamiltonian, for example, the Hamiltonian of two-dimensional
isotropic harmonic oscillator, $ \hat{H}(\hat{x},\,\hat{p}) =
\frac{1}{2m}(\hat{p}_x^2 + \hat{p}_y^2) +
\frac{m\,\omega^2}{2}(\hat{x}^2 + \hat{y}^2)$, in order to maintain
the physical meaning of $\hat{a}$, $\hat{b}$, $\hat{a}^{\dag}$ and
$\hat{b}^{\dag}$\,, i.e., to keep the relations among $(\hat{a},
\hat{b}, \hat{a}^{\dag}, \hat{b}^{\dag})$ and $(\hat{x}, \hat{y},
\hat{p}_x, \hat{p}_y)$ having the same formulations as the ones in
commutative space, one must let the model parameters $m$ and
$\omega$ satisfy a relation $m^2 \omega^2 = \mu /\nu$, which is
equivalent to propose a direct proportionality between the
noncommutative parameters $\mu$ and $\nu$\,. In our opinion, the
parameters $\mu$ and $\nu$ reflect the intrinsic noncommutativity
between positions and momenta respectively, (like as the Planck
constant encodes the non-commutativity of position and momentum,)
which should be independent on the concrete physical model.

We would like also to emphasize that the coherent state
representation and the squeezed state representation in the
noncommutative phase space are constructed in this Letter. Besides
this, we can also construct continuum entangled state representation
for the NCQM. These constructed representations will certainly play
important roles in studying physical problems on the noncommutative
phase space. The related result will be reported in a following
separated paper.

\end{document}